\documentclass[article]{IEEEtran}
\IEEEoverridecommandlockouts
\usepackage{cite}
\usepackage{amsmath,amssymb,amsfonts}
\usepackage{algorithmic}
\usepackage{graphicx}
\usepackage{textcomp}
\usepackage{psfrag}
\usepackage{algorithm2e}
\usepackage{url}
\usepackage{xcolor}
\def\BibTeX{{\rm B\kern-.05em{\sc i\kern-.025em b}\kern-.08em
    T\kern-.1667em\lower.7ex\hbox{E}\kern-.125emX}}

    \newcommand{\CCH}[1]{{#1}}
    
    \usepackage[absolute]{textpos}
    \newcommand{\copyrightstatement}{
    \begin{textblock}{15}(0.5,0.3)    
         \noindent
         \centering
         \textblockcolour{white}
         \footnotesize
         \copyright 2023 IEEE. Personal use of this material is permitted. Permission from IEEE must be obtained for all other uses, in any current or future media, including reprinting/republishing this material for advertising or promotional purposes, creating new collective works, for resale or redistribution to servers or lists, or reuse of any copyrighted component of this work in other works
    \end{textblock}
}

\begin{document}

\title{Extended Signaling Methods for Reduced Video Decoder Power Consumption Using Green Metadata\vspace{-.5cm}}

\copyrightstatement

\author{\IEEEauthorblockN{Christian Herglotz\textsuperscript{1}, Matthias Kr\"anzler\textsuperscript{1}, Xixue Chu\textsuperscript{1}, Edouard Fran\c{c}ois\textsuperscript{2}, Yong He\textsuperscript{3}, Andr\'e Kaup\textsuperscript{1}}\\
\IEEEauthorblockA{\textit{\textsuperscript{1}Multimedia Communications and Signal Processing} \\
\textit{Friedrich-Alexander University Erlangen-N\"urnberg}, Erlangen, Germany \\
\textit{\textsuperscript{2}Interdigital}, Rennes, France} \\
\textit{\textsuperscript{3}Qualcomm Techhnologies, Inc.}, San Diego, CA, USA \\
\{christian.herglotz, matthias.kraenzler, xixue.chu, andre.kaup\}@fau.de, edouard.francois@interdigital.com, yonghe@qti.qualcomm.com \vspace{-1.2cm}}

\maketitle

\begin{abstract}
 In this paper, we discuss one aspect of the latest MPEG standard edition on energy-efficient media consumption, also known as Green Metadata (ISO/IEC 232001-11), which is the interactive signaling for remote decoder-power reduction for peer-to-peer video conferencing. In this scenario, the receiver of a video, e.g., a battery-driven portable device, can send a dedicated request to the sender which asks for a video bitstream representation that is less complex to decode and process. Consequently, the receiver saves energy and extends operating times. We provide an overview on latest studies from the literature dealing with energy-saving aspects, which motivate the \CCH{extension of the legacy Green Metadata standard. Furthermore, we explain the newly introduced syntax elements and verify their effectiveness by performing dedicated} experiments. We show that the integration \CCH{of these syntax elements can lead to dynamic energy savings of up to $90\%$ for software video decoding and $80\%$ for hardware video decoding, respectively. } 
 
\end{abstract}


\vspace{-.4cm}

\section{Introduction}

\vspace{-.1cm}

In the recent years, we have witnessed an enormous growth in online video communication services. Nowadays, more than $75\%$ of the total Internet traffic constitutes video data \CCH{\cite{cisco20}}. Billions of users worldwide regularly utilize video communication. Studies estimate that roughly $1\%$ of global greenhouse gas emissions are caused by online video applications \cite{ShiftFull19}. Hence, solutions for the energy-efficient use of this technology are an important contribution towards reducing carbon emissions and fighting climate change.

As it is well known for decades that reducing the energy consumption in information and communication technology (ICT) is important, various papers presented solutions to assess and reduce the energy consumption of video communication devices. The power consumption of smartphones was investigated during video playback in online and offline scenarios \cite{Carroll13}. Dedicated power consumption models were derived, which could be exploited to achieve power savings while keeping the visual quality \cite{Li12,Herglotz19b}. In a similar direction, a power model for video streaming solutions was proposed for laptops and desktop PCs allowing to accurately estimate the power demand depending on high-level video parameters \cite{Herglotz22a}. Furthermore, many studies target the complexity reduction of encoders and decoders \CCH{for different codecs such as versatile video coding (VVC) or high-efficiency video coding (HEVC)}, which ultimately lead to energy savings during runtime \cite{Kraenzler22,Mallikarachchi20,Correa18}. 

\begin{figure}
\centering
\psfrag{S}[c][c]{Sender}
\psfrag{R}[c][c]{Receiver}
\psfrag{E}[r][r]{\textbf{!!}}
\psfrag{I}[c][c]{P2P-Connection}
\psfrag{D}[c][c]{DOR-Req}
\includegraphics[width=0.36\textwidth]{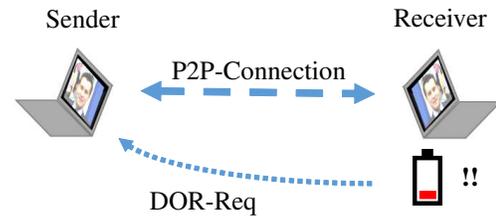}
\vspace{-0.7cm}
\caption{Interactive Signaling for remote decoder-power reduction in a P2P videoconferencing scenario as defined in the Green Metadata standard \cite{GREEN-MPEG}. }
\label{fig:DORR}
\vspace{-0.7cm}
\end{figure}

In a similar direction, the Moving Pictures Experts Group (MPEG) started an activity to define standards on tools and methods to reduce the energy consumption of video communication technologies. This activity formed the Green Metadata standard (referred to as ISO/IEC 232001-11), whose first edition was finalized in July 2015 \cite{GREEN-MPEG,Fernandes15}. In this standard, methods associated with metadata signalling mechanisms were proposed to reduce the energy consumption of various devices in video communication. Concerning the sender, it was proposed to perform low-power encoding \CCH{by reducing the visual quality of the compressed video. To allow quality-recovery at the receiver side, additional quality information such as, e.g., the peak signal-to-noise ratio (PSNR) or the structural similarity (SSIM) of the reconstructed video with respect to the original video could be sent as metadata \cite{Fernandes15}, which helps in guided image enhancement methods \cite{Wen13}. In addition}, it was proposed to signal complexity metrics as metadata to the receiver that allow early decoder complexity estimates. These estimates can be used to control the frequency of the receiver's central processing unit (CPU) \cite{Benmoussa16}. Furthermore, it was proposed to send brightness statistics for adapting the backlight brightness in liquid-crystal displays (LCDs), which can be exploited to reduce the power consumption \cite{Fernandes19}. Finally, a technique called ``interactive signaling for remote decoder-power reduction'' was proposed, in which the receiver sends a request to the sender asking for reducing the processing complexity of decoding. In this paper, we review this method, explain the new types of requests which were adopted in the latest version of the Green Metadata standard, and show that this new syntax allows for \CCH{receiver-side} decoding energy savings of up to \CCH{$90\%$}. 

This paper is organized as follows. In Section~\ref{sec:review}, we will briefly explain the legacy definition of the interactive signaling and motivate the adoption of new request types. Then, Section~\ref{sec:syntax} will present the newly adopted syntax elements and explain their meaning. Subsequently, Section~\ref{sec:exp} presents experiments that \CCH{show the potential energy savings of all syntax elements}. Finally, Section~\ref{sec:concl} concludes this paper. 


\section{Interactive Signaling for Remote Decoder-Power Reduction}
\label{sec:review}
\vspace{-.1cm}

In the original Green Metadata standard, interactive signaling for remote decoder-power reduction was proposed with the target of reducing the power consumption in peer-to-peer (P2P) videoconferencing applications \cite{GREEN-MPEG}. Fig.~\ref{fig:DORR} illustrates the main functionality of this method. In a P2P video conference, one side is defined as the sender (left) and the other side as the receiver (right). 
At a certain moment, the receiver desires to reduce its power consumption because of, e.g., a low battery level. 
The legacy Green Metadata allows receiver-side power reductions by requesting \CCH{bit streams with reduced decoding CPU operations, i.e., reduced decoding complexity.} 

Reducing complexity can decrease the receiver power in two ways. First, a reduced number of operations directly \CCH{results in} a reduced energy consumption \cite{Herglotz18}. Second, the CPU frequency can be reduced while still keeping real-time restrictions using dynamic voltage and frequency scaling (DVFS), which leads to additional power savings \cite{Tseng10}. In Green Metadata, the main idea is that the encoder at the sender constructs a bitstream that requires fewer decoder operations. This request is called ``decoder operations reduction request (DOR-req)''.

To implement the DOR-req., a corresponding syntax element was proposed requesting a reduction in decoder operations, which was called \texttt{dec\_ops\_reduction\_req}. This value, coded in a signed integer 8-bit representation, can be transformed to a percentage change of decoder operations $c\in [-100\%,100\% ]$ and is calculated by 
\begin{equation}
c = 100\% \ \frac{\texttt{dec\_ops\_reduction\_req}}{128}.
\label{eq:DORreq}
\end{equation}
According to the requested percentage change, the encoder at the sender can adapt its encoding parameters to meet the requested requirements. Positive performance changes ($c>0\%$) are allowed because in practice, battery-driven devices could be plugged to a power supply such that a reduced power consumption would not be needed anymore. 

\CCH{It is worth mentioning that in principle, a request for a large reduction of the decoding operations ($<-50\%$) cannot be inverted by a single request to increase the operations. The reason is that the inverse of halving the operations is doubling the operations, which corresponds to an increase of $100\%$. In practice, however, this can be solved by sending multiple requests for a positive change successively. }

So far, no work has been done or proposed on the practical application of the DOR-req. However, in research independent from the Green Metadata, several approaches were proposed in the literature that achieve decoder complexity savings using encoder parameters. 
For example, it was proposed to include the decoding complexity in the rate-distortion optimization process such that low-complex decoding tools are chosen \cite{Mallikarachchi20, Correa18, Herglotz20b}. \CCH{Also, common encoder implementations provide a tuning targeting fast or low-energy decoding, which causes energy savings (e.g., \texttt{fastdecode} tuning for x265 \cite{x265} and the \texttt{lowDecEnergy} configuration in VVenC \cite{Kraenzler22a}).} Unfortunately, all these solutions mainly target software decoders and no energy reductions were reported for hardware decoders, which are usually used on portable devices. \CCH{However, for hardware decoders,} it is reported that significant energy can be saved using spatial and temporal scaling or bitrate adaptions \cite{Herglotz19b,Herglotz23a,m57977}. 
\CCH{In this paper, we report decoding energy savings for both hardware and software decoders for all syntax elements included in the latest Green Metadata standard.}


\vspace{-.5cm}

\section{Green Metadata Syntax v3}
\label{sec:syntax}
\vspace{-.2cm}

\begin{table*}[h!t]
\caption{Syntax elements for DOR-Reqs. and corresponding encoder configuration flags.   }
\centering
\vspace{-.3cm}
\begin{tabular}{l|l|l|r|l}
\hline
Syntax element & Description  & Range & Bits& Encoder flag\\ 
\hline
\texttt{dec\_ops\_reduction\_req} & Change in decoder operations & $[-62,64]$ & $6$ & x265: \texttt{--derdo} \\
\texttt{disable\_loop\_filters} & Enable/disable loop filters & $\{0,1\}$ & $1$ & x265: \texttt{--no-dbf, --no-sao}\\
\texttt{disable\_bi\_iprediction} & Enable/disable bi-prediction & $\{0,1\}$ & $1$ & x265: \texttt{--bframes 0}\\
\texttt{disable\_intra\_in\_B} & Enable/disable intra prediction in B-frames & $\{0,1\}$ & $1$ & x265: \texttt{--no-b-intra}\\
\texttt{disable\_fracpel\_filtering} & Enable/disable fractional-pel filtering operations & $\{0,1\}$ & $1$ & x265: forbid fractional pel filterings	\\
\texttt{pic\_width\_in\_luma\_samples} & Desired horizontal resolution & $[0,16383]$ & $14$ & ffmpeg: \texttt{-vf scale}\\
\texttt{pic\_height\_in\_luma\_samples} & Desired vertical resolution & $[0,16383]$ & $14$& ffmpeg: \texttt{-vf scale}\\
\texttt{frames\_per\_second} & Desired frame rate in frames per second (fps) & $[0,1024]$& $10$ & ffmpeg: \texttt{-r} \\ \hline
\end{tabular}
\label{tab:syntax}
\vspace{-.6cm}
\end{table*}

\CCH{In the third edition of the Green Metadata standard \cite{GREEN-MPEG3rd}, the} legacy interactive signaling procedure \CCH{is extended} with further syntax elements which explicitly target the \CCH{energy reduction methods} mentioned above.
 As the receiver is aware of the decoder implementation it is using, it can request the ideal encoder configuration such that the receiver's power consumption is reduced maximally while keeping a decent visual quality. 

The \CCH{corresponding} syntax elements 
are shown in Table~\ref{tab:syntax}. 
The legacy request for decoder operations reduction \CCH{is redefined} with a modified reduction range (\texttt{dec\_ops\_reduction\_req}). While in the original implementation, a range of $[-100, 100]$ was allowed, the range \CCH{is now restricted }to $[-62,64]$. The reason is that an energy reduction of $100\%$ is infeasible in practice and that maximum reported savings were \CCH{in the redefined} range \cite{Herglotz20b,Kraenzler22}. 
As $6$ bits are used for signaling resulting in $64$ available values, only even percentage numbers can be chosen. 

Concerning coding tools, the receiver can request to enable or disable loop filters (\texttt{disable\_loop\_filters}), bi-prediction (\texttt{disable\_bi\_prediction}), intra-prediction in B-frames (\texttt{disable\_intra\_in\_B}), or fractional-pel filtering operations (\texttt{disable\_fracpel\_filtering}). 
For signaling, each tool is assigned to one bit. \CCH{To keep the syntax independent of a specific standard}, the loop filter is not specified. \CCH{It is up to the encoder to decide whether one or multiple loop filters are disabled.} Depending on the \CCH{used standard}, the loop filter could, e.g., be the deblocking filter (DBF), sample adaptive offset (SAO), or the adaptive loop filter (ALF). \CCH{We do not consider luma mapping with chroma scaling (LMCS), which is sometimes referred to as a loop filter in VVC \cite{Bross21}, because unlike other loop filters, it performs processing steps in the core decoding loop and not only before saving a frame in the decoded picture buffer.}

Finally, high-level video parameters can be \CCH{requested, namely the} spatial resolution in terms of picture width (\texttt{pic\_width\_in\_luma\_samples}) and height (\texttt{pic\_height\_in\_luma\_samples}) in luma samples as well as the temporal resolution in terms of the frame rate (\texttt{frames\_per\_second}). The numbers of bits are chosen in such a way that all resolutions and frame rates used in modern video formats are covered. \CCH{In the next section, we will report actual energy savings achieved using these newly defined syntax elements. }

\vspace{-.5cm}

\section{Experiments}
\label{sec:exp}
\vspace{-.2cm}

In this section, we present experiments \CCH{showing the effectiveness of these syntax elements. We cover hardware and software decoding of HEVC coded sequences, software decoding of VVC with different software implementations, and power measurement results of a fully functional P2P video communication setup.}


\subsection{Energy Savings for HEVC}
\label{sec:Rock}
\vspace{-.1cm}
\CCH{We construct dedicated video bit streams for each of the syntax elements mentioned in the last section as follows. For \texttt{dec\_ops\_reduction\_req}, we }take the x265 encoder with a decoding-energy-rate-distortion optimization (DERDO) add-on that was presented in \cite{Herglotz20b}. In this implementation, the rate-distortion optimization process is extended by considering the expected decoding energy. The coding costs are minimized 
\begin{equation}
\min J= D + \lambda_\mathrm{R}R + \lambda_\mathrm{E}E, 
\end{equation} 
where $D$ is the distortion, $R$ the rate, $E$ the decoding energy, and $\lambda_\mathrm{R}$ and $\lambda_\mathrm{E}$ two Lagrange multipliers indicating the desired trade-off between distortion, rate, and decoding energy. The decoding energy $E$ is estimated by a linear model 
\begin{equation}
E=\sum_{i=1}^N n_i\cdot e_i, 
\end{equation}
where for each \CCH{encoder} decision, the corresponding expected decoding energy $E$ is estimated by summing over a set of $N$ coding tools. Each coding tool $i$ can occur $n_i$ times and consumes $e_i$ joules of energy during decoding. 

\CCH{To test the impact of removing fractional pel filtering, we use the same DERDO implementation but 
set the energy parameter for fractional pel filtering } to a large number ($e_\mathrm{fracpel}=2^{16}$), such that the DERDO process avoids choosing it for coding. All the other $e_i$ are set to zero such that they have no further influence on encoder decisions. 

\CCH{The loop filters deblocking filter (DBF) and sample adaptive offset (SAO) are disabled by the corresponding flags available in the x265 encoder (\texttt{--no-deblock} and \texttt{--no-sao}) \cite{x265}. A similar flag is available to disallow intra prediction in B frames (\texttt{--no-b-intra}). Bi-prediction is disabled by setting the rate of B-frames to zero (\texttt{--bframes 0}). Temporal and spatial scaling is performed using FFmpeg \cite{FFmpeg} filters before compression. Temporal scaling, i.e., the reduction of frames per second (fps), is implemented by frame dropping, spatial scaling is implemented by bilinear filtering. The encoder settings are summarized in the last column of Table~\ref{tab:syntax}.}

\begin{table}[t]
\centering
\caption{\CCH{Measured energy savings (positive values mean a lower energy consumption) and BDR values for hardware and software decoding on the evaluation board.  }}
\label{tab:rock}
\vspace{-.3cm}
\resizebox{.5\textwidth}{!}{ 
\begin{tabular}{r|r|r|r|r|r|r}
\hline
 & \multicolumn{2}{c|}{Software} & \multicolumn{2}{c|}{Hardware} & \multicolumn{2}{c}{BDR}\\
 & Class B & Class E & Class B & Class E & Class B & Class E\\ \hline
derdo & $35.76\%$ & $28.21\%$ & $3.70\%$ & $1.86\%$ & $56.43\%$ & $37.24\%$ \\ 
no DBF & $16.64\%$ & $7.96\%$ & $2.59\%$ & $-0.40\%$ & $20.47\%$ & $18.13\%$ \\ 
no Sao & $6.36\%$ & $0.94\%$ & $1.81\%$ & $0.20\%$ & $12.57\%$ & $9.64\%$ \\ 
no Bi & $16.57\%$ & $32.97\%$ & $6.88\%$ & $7.48\%$ & $78.97\%$ & $81.03\%$ \\ 
no Intra In B & $3.79\%$ & $0.03\%$ & $-0.91\%$ & $0.43\%$ & $14.05\%$ & $10.36\%$ \\ 
no fracpel & $40.28\%$ & $24.00\%$ & $7.61\%$ & $2.11\%$ & n/a & $130.74\%$ \\ \hline
Res: 720p & $58.32\%$ & $0\%$ & $47.27\%$ & $0\%$ & $72.65\%$ & $0\%$ \\ 
Res: 540p & $77.14\%$ & $48.77\%$ & $64.55\%$ & $34.82\%$ & n/a & $46.44\%$ \\ 
Res: 360p & $89.64\%$ & $77.92\%$ & $78.21\%$ & $61.95\%$ & n/a & n/a \\ \hline
half fps & $43.07\%$ & $43.76\%$ & $43.71\%$ & $44.76\%$ & n/a & $38.06\%$ \\ 
third fps & $58.69\%$ & $60.19\%$ & $58.27\%$ & $60.06\%$ & n/a & $95.27\%$ \\ 
quarter fps & $66.96\%$ & $68.43\%$ & $66.04\%$ & $67.67\%$ & n/a & n/a \\ 
\hline
\end{tabular}}
\vspace{-.5cm}
\end{table}

\CCH{As input sequences, we choose sequences from the JVET common test conditions \cite{JVET_N1010}: five sequences from class B (HD resolution) and three sequences from class E (720p resolution), whose content is  comparable to video conferencing applications (persons talking in front of a static background). The sequences are encoded with constant rate factors (crf) 18, 23, 28, and 33. }

\CCH{For our measurements, we use a Rock 5B board \cite{rock5b} with a Rockchip RK3588 System-on-Chip (SoC), which supports HEVC hardware decoding. The CPU is a quad-core ARM Cortex-A76 MPCore and a quad-core ARM Cortex-A55 MPCore. 
The operating system is Ubuntu. We measure the energy consumption through the main power supply of the board using an external power meter (an LMG611 by ZES Zimmer). For both software and hardware decoding, we use FFmpeg and configure the decoding process accordingly. We report results for the dynamic energy consumption, i.e., we neglect the static, idle energy consumption of the board. A Student's t-test is performed to ensure statistical validity of the measurement. More detailed information can be found in \cite{Herglotz18}. }

\CCH{Table~\ref{tab:rock} summarizes the measurement results. Each row reports energy savings for a single syntax element as positive values and the corresponding rate-distortion performance in terms of the Bj{\o}ntegaard-Delta rate (BDR) calculated using Akima interpolation \cite{Herglotz22b}. For spatial downscaling to 720p of class E, the values are zero because this is the native resolution of the sequences. `n/a' means that for at least one of the sequences, the rate-distortion curves showed no overlap in the PSNR domain such that BDR values cannot be calculated. }

\CCH{Concerning software decoding, we find that almost all methods lead to significant energy savings. Highest energy savings are observed at the lowest resolution 360p (up to $90\%$ and $80\%$ for class B and E, respectively), which can be expected because the number of pixels to be decoded is reduced by a factor of nine. On the other hand, the visual quality is also reduced to an extent in which the BDR cannot be calculated anymore (n/a). Lowest savings are reported for \texttt{--no-b-intra}. Furthermore, we can see that savings highly depend on the content. For example, disabling bi-prediction leads to more energy savings in class E than in class B, for disabling DBF, we can observe the inverse behavior. Taking the BDR into account, we find that \texttt{derdo}, disabling DBF, and frame rate reduction lead to good compromises between energy reduction and compression performance for software decoding.  }

\CCH{With regards to hardware decoding, we can find that energy savings differ significantly from software energy savings. When disregarding spatial or temporal scaling, notable energy savings above $5\%$ can only be observed for disabling bi-prediction and fractional pel interpolations. In some cases, even a slightly higher energy consumption was observed ($<1\%$), which could be caused by measurement noise. Still, we find that strong energy savings are obtained by temporal and spatial scaling of the videos. For this, between $30\%$ and $70\%$ of energy savings are observed. }

\begin{table}[t]
\centering
\caption{Measured energy and time savings for VVC software decoding when disabling a tool (first column). Tests were performed for single-thread (ST) and multi-thread (MT) execution. The results indicate energy/time savings \CCH{as well as BDR increases} with a tool switched off.}
\vspace{-.3cm}
\resizebox{.5\textwidth}{!}{ 
\begin{tabular}{r|r|r|r|r}
\hline
Hardware	 $\rightarrow$	& Intel-i7 & Intel-i7& Intel-i7   &  \\
Tool $\downarrow$ Software$\rightarrow$ & VVdeC (ST) & VVdeC (MT) & VTM (ST) & \CCH{BDR}\\
\hline
DBF & $13.03\%$  & $5.88\%$  & $10.47\%$    & \CCH{$0.73\%$} \\
SAO & $2.01\%$  & $0.32\%$  & $0.49\%$   & \CCH{$0.19\%$} \\
ALF & $14.14\%$  & $12.85\%$  & $7.08\%$    &\CCH{ $5.79\%$} \\
\hline
Bi-pred. & $3.74\%$  & $1.82\%$  & $1.09\%$  & \CCH{$3.94\%$} \\
\hline
\end{tabular}}
\label{tab:toolSavings}
\vspace{-.6cm}
\end{table}

\vspace{-.4cm}

\subsection{Energy Savings for VVC}
\label{sec:Intel}
\vspace{-.2cm}

\CCH{To show that the syntax elements can also be helpful for other codecs than HEVC, we tested the performance of selected syntax elements on VVC decoding. To this end, we performed} processing energy measurements for two implementations of a VVC decoder on a desktop PC. The two decoder implementations are VTM-11.0 \cite{VTM} and VVdeC v1.0.0 \CCH{\cite{Wieckowski20}}. The desktop PC is an Intel i7-8700 CPU. Measurements are performed for all HD sequences of the JVET common test conditions (CTC) \cite{JVET_T2010}. \CCH{The energy is measured using running average power limit (RAPL) \cite{David10}, such that only the power consumption of the Intel core is considered. Similar to the HEVC measurements, we focus on the dynamic energy.} We report mean relative energy savings over all sequences and the standard four CTC \CCH{quantization parameters (QPs). For VVC}, we focus on loop filters \CCH{and bi-prediction. The former can be switched on and off using encoder settings, the bi-prediction is disabled by comparing the lowdelay\_P with the lowdelay\_B configuration. Results are summarized in Table~\ref{tab:toolSavings}.}

 The table lists savings for three different loop filters: \CCH{DBF, SAO, and the adaptive loop filter (ALF),} which is only available in VVC. We can observe that the savings are highly variable with respect to the \CCH{processing configuration}. While switching off DBF results in almost $15\%$ energy savings in VVdeC single-thread processing, the savings are less than $6\%$ for multi-threading. Similar observations hold for ALF, where savings range from $7\%$ up to $15\%$, depending on the software. Similar to HEVC, SAO has little impact on the energy consumption (always below $3\%$ savings). \CCH{Furthermore, we report corresponding increases in bitrate in terms BDR, which shows that the DBF provides the best trade-off between compression efficiency loss and energy efficiency improvement for VVC. }

\begin{figure}
\centering
\psfrag{S}[c][c]{Sender}
\psfrag{R}[c][c]{Receiver}
\psfrag{I}[c][c]{P2P-Connection}
\psfrag{P}[r][r]{Power}
\psfrag{M}[r][r]{Meter}
\includegraphics[width=0.35\textwidth]{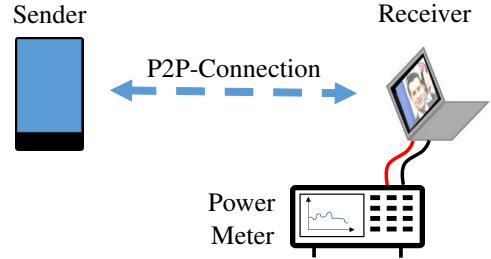}
\vspace{-.5cm}
\caption{Power measurement setup for the Laptop. }
\label{fig:setup}
\end{figure}
\begin{table}[t]
\centering
\caption{Measured power savings by spatiotemporal downsampling with respect to the default values (bold) for the web application \CCH{(hardware decoding). $0\,$fps corresponds to a still image.  } }
\label{tab:resXixue}
\vspace{-.3cm}
\begin{tabular}{r|r||r|r}
\hline
Frame rate scaling & Savings & Resolution scaling & Savings \\
\hline
$\mathbf{30}\,$ \textbf{fps @} $\mathbf{1080}$\textbf{p} & $0\%$ & $\mathbf{1080}$\textbf{p} \textbf{@} $\mathbf{30}\,$ \textbf{fps} & $0\%$\\ 
$20\,$ fps @ $1080$p & $4.97\%$ & $720$p @ $30\,$ fps & $1.65\%$\\ 
$10\,$ fps @ $1080$p & $17.03\%$ & $540$p @ $30\,$ fps & $3.55\%$\\ 
$0\,$ fps @ $1080$p & $20.83\%$ & $360$p @ $30\,$ fps & $9.11\%$\\ 
\hline
\end{tabular}
\vspace{-.6cm}
\end{table}

\vspace{-.4cm}

\subsection{P2P Video Conferencing}
\label{sec:P2P}
\vspace{-.2cm}

\CCH{Finally,} we perform a dedicated experiment on an actual conferencing application \CCH{to validate energy savings in a realistic P2P scenario}. The P2P application is based on WebRTC \CCH{\cite{WebRTC}} and  Firebase \CCH{\cite{Firebase}}, it uses the internal hardware for video decoding\CCH{, and employs H.264 as video codec}. 

The measurement platform as shown in Fig.~\ref{fig:setup} consists of a laptop as a receiver, an external power meter (ZES Zimmer LMG95), and a smartphone as a remote sender. The laptop is a Dell Vostro 5590 equipped with an \CCH{Intel Core i7-10510U@1.8GHz CPU} and a liquid crystal display of $15.6\,$inches, $1920\times 1080$ resolution. The operating system is Windows 10. We measure the power consumption through the main power supply of the laptop. During measurements, the battery is fully loaded to ensure that battery charging does not interfere our measurements. The remote sender is a Samsung A20e. 
To have full control of the video content, we choose the `Johnny' sequence from the JVET common test conditions \cite{JVET_T2010}, which has content similar to videoconferencing content (a sitting and slightly moving person in front of a static background). We also verify results on a real video taken from the sender's camera, where the user is talking to a virtual peer on the remote side.

 The user of the application can choose different frame rates and video resolutions \CCH{ corresponding to the last three syntax elements in Table~\ref{tab:syntax}. We did not test further syntax elements because we relied on available encoder configurations in the WebRTC framework. } The default values are $30\,$fps and $1920\times 1080$, respectively\CCH{, at a bitrate of $1{,}500\,$kbps}. \CCH{Note that instead of temporal downscaling, the decoder could also choose to discard higher temporal layers of the sequence. In practice, however, the observed energy savings would be smaller because the bitrate and hence the power consumption of the receiver module would not be affected \cite{Herglotz20}. Also, some devices might not support temporal scalability. } 
 
 Relative power savings when reducing the frame rate or the resolution are listed in Table~\ref{tab:resXixue}. \CCH{In contrast to the decoder experiments above, we report full power savings including peripheral components of the laptop, e.g., the screen, such that we expect lower savings. } 
 We can see that for both downscaling algorithms, significant power savings can be reached. Apparently, potential savings are larger for frame rate reductions (up to $20\%$ for still pictures) than for spatial scaling (up to $10\%$). 


\section{Conclusion}
\label{sec:concl}
\vspace{-.2cm}
In this paper, we have presented the latest update on the Green Metadata standard concerning interactive signaling for remote decoder-power reduction. \CCH{For the newly introduced syntax elements,} we presented a set of dedicated experiments with results on energy and power savings. \CCH{First}, we showed that \CCH{achievable energy savings highly depend on the decoder's hardware and software.}  For software decoding, we reached \CCH{dynamic decoding} energy savings up to $90\%$. For hardware decoders, we \CCH{reached dynamic energy} savings up to $80\%$. \CCH{On an actual conferencing app on a laptop, we showed that these decoding energy savings lead to up $20\%$ of power savings if the device's total power consumption is considered. }

In future work, the request messages could be generated by an automated monitoring script on the receiver side. This script could choose the power-reduction tool depending on the battery status. Furthermore, other tools for decoder-power reduction such as codec changes or codec-specific requests could be investigated. \CCH{Also, a framewise tool switching as proposed in \cite{Kraenzler22a} could be included as syntax element.} Finally, the application of similar concepts for multicast and broadcast scenarios could be investigated.

%

	
\bibliographystyle{IEEEbib}

\end{document}